\documentclass[10pt]{article}
\usepackage{amssymb}

\begin{document}
\title{Topological mass in seven dimensions and dualities in four dimensions}
\author{M. Betancourt and A. Khoudeir\footnote{E-mail:minerba@ula.ve and
adel@ula.ve}\\
{\it Centro de F\'isica Fundamental, Facultad de Ciencias,}\\ {\it Universidad de los Andes, M\'erida 5101, Venezuela}}

\maketitle

\begin{abstract}
The massive topologically and self dual theories en seven
dimensions are considered. The local duality between these
theories is established and the dimensional reduction lead to the
different dualities for massive antisymmetric fields in four
dimensions.
\end{abstract}
The topological Chern Simons terms have played an important role
in several physical models. For instance, they appear in the
eleven dimensional supergravity in a natural way {\cite{primera}}
and arise in the anomalies cancellation for gauge and string
theories. Also, they allow the formulation of genuine gauge theory
for gravity.{\cite{segunda}}. If they are included in the
conventional theories for odd dimension $D=4k-1$, it is possible
formulate massive theories which are compatible with gauge
invariance. Initially, this goal was formulated in three
dimensions provide gauge invariant theories for massive spin 1 and
spin 2 fields {\cite{tercera}}. These are called massive
topologically theories. Alternatively, other formulations describe
the same physical dynamics but in a non-gauge invariant way: the
self dual theories \cite{cuarta}\cite{quinta} Eventually, it was
established that they are essentially two ways for describing the
same physics\cite{sexta}, i.e., they are related by
duality\cite{septima}. Furthermore, the dual equivalence can be
shown from the hamiltonian framework\cite{octava}. The self dual
theories are gauged fixed versions of the topologically massive
theories. Also, there exist analogues dualities for antisymmetric
fields in four dimensions, \cite{novena},\cite{diez}, which
constitute alternative manners to describe massive scalar and
vectorial fields through of gauge invariant topological BF terms.

In this work, we will consider the dual equivalence between the
topologically massive and self dual theories in seven dimensions.
The basic field is a third order antisymmetric tensor. Recently,
it was recognized the importance of these theories, when the
eleven dimensional supergravity is dimensionally reduced in a
consistent way on $A_dS_7 {\otimes} S_4\cite{nueva}$. We perform
the Hamiltonian analysis in order to achieve the local canonical
duality between them. Besides, the duality can be inferred through
the existence of a first order master action. Finally, we will
make the dimensional reduction of the topologically massive and
self dual theories in seven dimensions, to obtain the several
dualities
for massive antisymmetric fields in four dimensions.\\
The topologically massive theory is described by the following action
\begin{equation}
I_{TM}=-\frac{1}{48}\int {d}^{7}xG_{mnqq}G^{mnpq}-\frac{\mu }{72}\int {d}^{7}
x\varepsilon ^{mnpqrst}C_{mnp}\partial _{q}C_{rst},
\end{equation}
where $C_{mnp}$ is a third order antisymmetric field,
$G_{mnpq}=\partial _{m}C_{npq}+\partial _{n}C_{pmq}+\partial
_{p}C_{mnq}+\partial _{q}C_{nmp}$ its field strength and $\mu $ is
a mass parameter. This action is invariant under gauge
transformations: $\delta {C_{mnp}}=\partial _{m}\Lambda
_{np}+\partial _{n}\Lambda _{pm}+\partial _{p}\Lambda _{mn}$. The
field equations for the topologically massive theory are
\begin{equation}
\partial _{m}G^{mnpq}=-\frac{\mu }{6}\varepsilon ^{mnpqrst}\partial
_{m}C_{rst}.
\end{equation}
On the other hand, the non gauge invariant self dual theory is also
formulated with a third order antisymmetric field ${\mathcal{C}_{mnp}}$. Its
action is
\begin{equation}
I_{SD}=\frac{1}{72\mu }\int {d}^{7}x\varepsilon ^{mnpqrst}{\mathcal{C}_{mnp}}%
\partial _{q}{\mathcal{C}_{rst}}-\frac{1}{12}\int {d}^{7}x{\mathcal{C}_{mnp}}%
{\mathcal{C}^{mnp}}.
\end{equation}
From this action, we have the following field equations
\begin{equation}
C^{mnp}=\frac{1}{6\mu }\varepsilon ^{mnpqrst}\partial _{q}C_{rst}.
\end{equation}
The thirty five components of $C_{mnp}$ can be decomposed in its transverse
and longitudinal parts in the following manner
\begin{equation}
{C}_{mnp}=\cases{C _{0ij}\equiv{b_{ij}}=b_{ij}^T+\rho_{i}b_{j}^T-\rho_{j}
b_{i}^T\qquad15=10+5\cr{C}_{ijk}=C_{ijk}^T+\rho_iC_{jk}^T+\rho_jC_{ki}^T+%
\rho_kC_{ij}^T\qquad20=10+10,}  \label{minerba}
\end{equation}
where $\rho _{i}=\partial _{i}/\sqrt{-\nabla ^{2}}$. We can show
that both theories propagate ten massive dynamical degrees of freedom
contained in $C_{ijk}^{T}$ and ${\mathcal{C}^T_{ijk}}$, i.e,
$(\Box -\mu ^{2})\left\{\begin{array}{c}C^{T}_{ijk}\\{\mathcal{C}^T_{ijk}}\end{array}\right\}$=0,
the same number of degrees of
freedom for a massless $C_{mnp}$ field and just the half of the massive $%
C_{mnp}$ field theory in seven dimensions. This suggest that they are dual
equivalent, similar to the three dimensional case. Indeed, we can see the
dual equivalence through the following master action
\begin{equation}
I_{M}=\frac{1}{36}\int {d}^{7}x\chi ^{mnp}{\mathcal{C}_{mnp}}-\frac{\mu }{72}%
\int {d}^{7}x\chi ^{mnp}C_{mnp}-\frac{1}{12}\int {d}^{7}x{\mathcal{C}_{mnp}}{%
\mathcal{C}_{mnp}},  \label{segunda}
\end{equation}

where
\begin{equation}
\chi^{mnp}\equiv\varepsilon^{mnpqrst}\partial_qC_{rst}.
\end{equation}
Making independent variations on ${\mathcal{C}_{mnp}}$ allow us determine ${%
\mathcal{C}_{mnp}}$ in terms of $C_{mnp}$
\begin{equation}  \label{primera}
{\mathcal{C}^{mnp}}=\frac{1}{6}\chi^{mnp}.
\end{equation}
Substituting ({\ref{primera}}) into ( {\ref{segunda}}) we find the
topologically massive action. Meanwhile, independent variations on $C_{mnp}$
tell us

\begin{equation}
\varepsilon^{mnpqrst}\partial_{q}({\mathcal{C}_{rst}}-\mu{C}_{rst})=0,
\end{equation}
which can be (locally) solved as $C_{mnp}=\frac{1}{\mu}{\mathcal{C}_{mnp}}%
+\partial_mA_{np}+\partial_nA_{pn}+\partial_pA_{mn}$ ($A_{mn}$
being gauge degrees of freedom, which do not affect the local
dynamics), then substituting into (\ref{segunda}) we obtain the
self dual action.\\ Now, let us look the dual equivalence from a
Hamiltonian framework. First, we consider the topologically
massive theory. The conjugate momenta are
\begin{equation}
\pi_{ijk}=\frac{\delta{\mathcal{L}_{TM}}}{\delta{\dot{C}_{ijk}}}=\frac{1}{6}(%
\dot{C}_{ijk}+\partial_iC_{j0k}+\partial_jC_{0ik}+\partial_kC_{i0j})-\frac{1%
}{72}\mu{\varepsilon}_{ijklmn}C_{lmn}
\end{equation}
and
\begin{equation}
\pi_{0ij}=\frac{\delta{\mathcal{L}_{TM}}}{\delta{\dot{C}_{0ij}}}=0\equiv{\phi%
}_{1ij}.
\end{equation}
This last is a primary constraint. The Hamiltonian is $H=\int{d}^6x({%
\mathcal{H}}+\chi_{1ij}\phi_{1ij})$ where ${\mathcal{H}}=\pi_{ijk}\dot{C}%
_{ijk}-{\mathcal{L}}$ is the Hamiltonian density and $\chi_{ij}$ is a
Lagrange multiplier. Preserving the primary constraint:$\dot{\phi}%
_{1ij}=\{\phi_{1ij}(x),H(y)\}\approx0$, we obtain a secondary
constraint
\begin{equation}
\phi_{2ij}=3\partial_k\pi_{kij}-\frac{1}{24}\mu\varepsilon_{ijklmn}%
\partial_kC_{lmn}\approx0.
\end{equation}
There is no more constraint, because $\dot{\phi}_{2ij}=\{\phi_{2ij}(x),H(y)\}%
\equiv0$. This set of constraints constitute a first class abelian gauge
algebra:
\begin{equation}
\{\phi_{\alpha{ij}},\phi_{\beta{kl}}\}=0\qquad{\alpha,\beta=1,2},
\end{equation}
which generate the gauge transformations for the topologically massive
theory. Finally, the extended Hamiltonian density is given by

\begin{eqnarray}
{\mathcal{H}_{TM}}&=&3\pi_{ijk}\pi_{ijk}+\frac{1}{48}\mu^2C_{ijk}C_{ijk} +%
\frac{1}{48}G_{ijkl}G_{ijkl}+{}  \nonumber \\
&&{}+ \frac{1}{12}\mu\varepsilon^{ijklmn}C_{lmn}\pi_{ijk}+\chi_{1ij}%
\phi_{1ij}+\chi_{2ij}\phi_{2ij},
\end{eqnarray}
where, $\chi_{1ij}$ and $\chi_{2ij}$ are the Lagrange multipliers,
associated to the constraints, $\phi_{1ij}$ and $\phi_{2ij}$,  respectively.
We can solve the constraints of the topologically massive theory,  after
defining a good fixing  gauge conditions, e.g., the Coulomb gauge $C_{0ij}=0$%
, $\partial_kC_{ijk}=0$, and show that only the ten components $C^T_{ijk}$
propagate  massively: $(\Box-\mu^2)C^T_{ijk}=0$.\newline
For the self dual theory, we have the following conjugate momenta
\begin{eqnarray}
\pi_{ijk}&=&\frac{\partial{\mathcal{L}}_{SD}}{\partial\dot{\mathcal{C}_{ijk}}}=\frac{1%
}{72\mu}\varepsilon^{ijklmn}{\mathcal{C}_{lmn}}, \\
\pi_{0ij}&=&\frac{\partial{\mathcal{L}}_{SD}}{\partial\dot{\mathcal{C}_{0ij}}}=0.
\end{eqnarray}
These are primary constraints
\begin{eqnarray}
\psi_{ijk}&=&\pi_{ijk}-\frac{1}{72}\mu{\varepsilon_{ijklmn}}{\mathcal{C}_{lmn}}\approx0
\\
\psi_{ij}&=&\pi_{0ij}\approx0.
\end{eqnarray}
The hamiltonian density is given by
\begin{equation}
{\mathcal{H}}=\frac{1}{12}{\mathcal{C}_{ijk}}{\mathcal{C}^{ijk}}-{\mathcal{C}_{0ij}}(6\partial_k\phi_{ijk}+\frac{%
1}{4}{\mathcal{C}_{0ij}}).
\end{equation}
Note that ${\mathcal{C}_{0ij}}$ plays the role of a Lagrange multiplier. Preserving the
primary constraint $\psi_{ij}$ ($\dot{\psi}_{ij}=\{H,\psi_{ij}\}\approx0$
with $H=\int{d}^6y[{\mathcal{H}}+\lambda_{ijk}\psi_{ijk}+\lambda_{ij}%
\psi_{ij}]$, $\lambda{^{\prime}}s$ being Lagrange multipliers), we obtain a
secondary constraint
\begin{equation}
\Theta_{ij}\equiv{-}6\partial_k\pi_{kij}-\frac{1}{2}{\mathcal{C}_{0ij}}\approx0,
\end{equation}
while the preservation of the other primary constraint, lead to the
determination of the Lagrange multiplier $\lambda_{ijk}$:
\begin{equation}
\lambda_{ijk}=\frac{1}{6}\mu{\varepsilon_{ijklmn}{\mathcal{C}_{lmn}}}-(%
\partial_i{\mathcal{C}_{0jk}}+\partial_j{\mathcal{C}_{0ki}}+\partial_k{\mathcal{C}_{0ij}}).
\end{equation}
Preserving the secondary constraint, we find the value of the Lagrange
Multiplier $\lambda_{ij}$
\begin{equation}
\lambda_{ij}=\partial_k{\mathcal{C}_{kij}}.
\end{equation}
The algebra of the constraints $\psi_{ij}$, $\psi_{ijk}$, $\Theta_{ij}$ is
second class, reflecting the non existence of gauge invariance in the
self dual action:
\begin{eqnarray}
\{\psi_{ijk},\psi_{lmn}\}&=&\frac{1}{{36}\mu}\varepsilon^{ijklmn}%
\delta^6(x-y), \\
\{\Theta_{ij},\psi_{klm}\}&=&\frac{1}{72\mu}\varepsilon^{ijklmn}\partial_n%
\delta^6(x-y), \\
\{\Theta_{ij},\psi_{lm}\}&=&\frac{1}{12}\delta^{ij}_{lm}\delta^6(x-y).
\end{eqnarray}
Obviously, we can solve the constraints, for instance, making use of
transverse-longitudinal decomposition given by eq.(\ref{minerba}). To show
the dynamical propagation of ${\mathcal{C}^T_{ijk}}$: $(\Box -\mu^2){%
\mathcal{C}^T_{ijk}}=0$.\newline
The Hamiltonian density of the self dual theory can be written down (after
redefining ${\mathcal{C}_{ijk}}\Rightarrow{\mu}C_{ijk}$) as
\begin{equation}
{\mathcal{H}_{SD}}=\frac{1}{12}\mu^2C_{ijk}C_{ijk}+\frac{1}{(12)^2}%
(\varepsilon^{ijklmn} \partial_kC_{lmn})^2.
\end{equation}
Having found the Hamiltonians of the massive topologically and
Self Dual Theories, we can establish the following relationship
between them
\begin{eqnarray}
{\mathcal{H}_{TM}}&=&{\mathcal{H}_{SD}}+3\psi_{ijk}(\psi_{ijk}+\frac{\mu}{18}%
\varepsilon_{ijklmn}C_{lmn})+{}  \nonumber \\
&&{}+\chi_{1ij}\phi_{1ij}+\chi_{2ij}\phi_{2ij}.
\end{eqnarray}
The two Hamiltonian densities are related by a specific combination of the second class constraint
$\psi_{ijk}$ of ${\mathcal{H}_{SD}}$.
This result shows the canonical duality equivalence. The self dual theory can be considered as a
gauge fixed version of the topologically massive theory. In fact, we observe that
$\frac{1}{3}\phi_{2ij}=\partial_k\psi_{kij}\approx0$ and $\pi_{0ij}\approx0$ can be
 considered as the first class constraints while $\Theta_{ij}$ and $\varepsilon^{ijklmn}\partial_k\psi_{lmn}\approx0$
 as the gauge fixing conditions.
It's worth recalling that this is a local equivalence. If we
consider a non trivial topological structure of space-time, then
we would expect that global equivalence is not hold exactly. In
this case, the partition functions will differ by a factor
associated with the topologically sectors not present in the space
of solutions of the self dual theory. This local duality
equivalence is valid even in the presence of sources.

Now, we will perform the dimensional reduction of the topologically massive
and self dual theories in seven dimensions to four dimensions. First, we
write the actions coupled to an arbitrary gravitational background:
\begin{eqnarray}
I_{TM}&=&-\frac{1}{48}\int{d}^{4+3}x\sqrt{-\hat{g}}\hat{g}^{MR}\hat{g}^{NS}%
\hat{g}^{PT}\hat{g}^{QV}\hat{G}_{MNPQ} \hat{G}_{RSTV}{}  \nonumber \\
&&{}-\frac{\mu}{72}\int{d}^{4+3}x\varepsilon^{MNPQRST}\hat{C}_{MNP}\hat{G}%
_{QRST}
\end{eqnarray}
and
\begin{eqnarray}
I_{SD}&=&\frac{1}{72\mu}\int{d}^{4+3}x\varepsilon^{MNPQRST}\hat{C}%
_{MNP}\partial_Q\hat{C}_{RST} {}  \nonumber \\
&&{}-\frac{1}{12}\int{d}^{4+3}x\sqrt{-g}\hat{g}^{MQ}\hat{g}^{NR}\hat{g}^{PS}%
\hat{C}_{MNP}\hat{C}_{QRS},
\end{eqnarray}
where $\hat{g}=det\hat{G}_{MN}$. We will compactify the seven dimensional
manifold on a four dimensional Mikownsky space time cross an internal
compact three dimensional manifold, i.e, ${\mathcal{M}_{7}}={\mathcal{M}%
_{4Mikownsky}}\otimes{\mathcal{M}_{3}}$:
\begin{equation}
ds^2=g_{MN}dx^Mdx^N=\eta_{mn}dx^mdx^n+g_{\alpha\beta}dx^\alpha{d}^\beta.
\end{equation}
The indices $m,n$ labels the Mikownskian indices and the greek indices $%
\alpha,\beta$ take values $\alpha,\beta=1,2,3$. The volume of the internal
manifold is taken to be the unit ($detg_{\alpha\beta}=1$). For the
antisymmetric third order field $\hat{C}_{MNP}$, we make the following
decomposition
\begin{equation}
\hat{C}_{mnp}=C_{mnp}(x)\quad {\hat{C}_{mn\alpha}}=B^\alpha_{mn}{(x)}\quad{
\hat{C}_{m\alpha\beta}}=\varepsilon_{\alpha\beta\gamma}A_m^{\gamma}(x) \quad{
C}_{\alpha\beta\gamma}=\varepsilon_{\alpha\beta\gamma}\phi(x).
\end{equation}
We drop all dependence of the internal coordinates. The respective strength
fields are
\begin{eqnarray}
\hat{G}_{mnpq}&=&G_{mnpq}=\partial_mC_{npq}+\partial_nC_{pmq}+%
\partial_mC_{npq}+\partial_qC_{nmp}, \\
\hat{G}_{mnp\alpha}&\equiv&H^{\alpha}_{mnp}=\partial_mB^{\alpha}_{np}+%
\partial_nB^{\alpha}_{pm}+\partial_pB^{\alpha}_{mn}, \\
\hat{G}_{mn\alpha\beta}&\equiv&\varepsilon_{\alpha\beta\gamma}F^{\gamma}_{mn}=%
\varepsilon_{\alpha\beta\gamma}
(\partial_mA^{\gamma}_n-\partial_nA^{\gamma}_m), \\
\hat{G}_{m\alpha\beta\gamma}&\equiv&\varepsilon_{\alpha\beta\gamma}\partial_m\phi.
\end{eqnarray}
With this ansatz, it is straightforward to make the dimensional
reduction process. For the topologically massive theory, we obtain
the following reduced action to four dimensions
\begin{eqnarray}
I_1&=&\int{d}^4x[-\frac{1}{4}F_{mn}^{\alpha}F^{mn\alpha}-\frac{1}{12}%
H_{mnp}^{\alpha}H^{mnp\alpha}-\frac{\mu}{4}\varepsilon^{mnpq}B_{mn}^{%
\alpha}F_{pq}^{alpha}]{}  \nonumber \\
&&{}+\int{d}^4x[-\frac{1}{48}G_{mnpq}G^{mnpq}-\frac{1}{2}\partial_m\phi%
\partial^m\phi-\frac{\mu}{6}\varepsilon^{mnpq}\phi{G}_{mnpq}].
\end{eqnarray}
The first integral is nothing but of a triplet of Cremmer-Scherk
actions\cite{decima} and as is well known, it describes in a gauge
invariant way, the dual equivalence between massive vector and
second order antisymmetric fields in four dimensions\cite{novena}.
This action allow us to obtain the non abelian Fredman-Townsend
\cite{decimaprimera} model from the self interaction
mechanism\cite{decimasegunda}. The second integral is just the
gauge invariant master action which allow us to show the dual
equivalence between massive scalar and a third order
antisymmetric fields in four dimensions\cite{diez}\\
On the other hand, the self dual action in seven dimensions is
reduced to four dimensions in the following way

\begin{eqnarray}
I_2 &=&\int {d}^{4}x[-\frac{1}{4}B_{mn}^{\alpha }B^{mn\alpha }-\frac{1}{2}%
A_{m}^{\alpha }A^{m\alpha }-\frac{1}{4\mu }\varepsilon ^{mnpq}B_{mn}^{\alpha
}F_{pq}^{\alpha }]{}  \nonumber \\
&&{}+\int {d}^{4}x[-\frac{1}{12}C_{mnp}C^{mnp}-\frac{1}{2}\phi ^{2}+\frac{1}{%
6\mu }\varepsilon ^{mnpq}\phi \partial _{m}C_{mnp}],
\end{eqnarray}
from which the dualities just commented above, is easily inferred
but in a non gauge invariant way. These dualities in four
dimensions are established in a local way. Global aspects of the
dual equivalences have been considered in \cite{novena} and
\cite{diez}.

 Summarizing, we have discussed some hamiltonian aspects of the
topologically massive and self dual theories in seven dimensions and we have
stated that they are dual equivalent from the Hamiltonian framework. Both
theories describe the same physical situation: the propagation of ten local
degrees of freedom in seven dimensions, contained in the purely transverse
part of $C_{MNP}\quad (C_{ijk}^{T})$. The self dual theory can be considered
as a fixed gauge version of the topologically massive theory. The dual
equivalence can be seen in a covariant way using the master action(6).
Since the constraints for higher antisymmetric field are reducible, these
theories deserve a full Hamiltonian treatment. It will be interesting the
BRST quantization of these theories. We have also obtained the different
local dualities between antisymmetric fields in four dimensions, after making the
dimensional reduction of the topologically massive and dual theories in
seven dimensions.
\section*{Acknowledgment}
The authors would like to thank to the Consejo de Desarrollo
Cient\'{\i}fico y Humanistico de la Universidad de los Andes for
institutional support under project C-1149-02-05-F.

\end{document}